\documentclass[sigconf]{acmart}

\AtBeginDocument{%
  }

\setcopyright{none}
\renewcommand\footnotetextcopyrightpermission[1]{}
\acmConference[Preprint]{Technical Report}{May 2026}{Indian Institute of Technology Bombay, India}
\acmBooktitle{Technical Report, May 2026, Indian Institute of Technology Bombay, India}

\begin{document}

\title{Operationalizing Cyber Threat Intelligence with GraphRAG}
\subtitle{Behavioral Knowledge Graphs for IOC-Resilient Threat Hunting Plan Generation}

\author{Atul Kabra, Prakhar Paliwal, and Manjesh K. Hanawal}
\email{\{atul.kabra, prakhar.paliwal, mhanawal\}@iitb.ac.in}
\affiliation{%
  \institution{MLiONS, Department of IEOR, IIT Bombay}
  \city{Mumbai}
  \state{Maharashtra}
  \country{India}
}

\renewcommand{\shortauthors}{Kabra}

\begin{abstract}
When a security researcher publishes a report on a cyberattack, detection engineers are supposed to turn it into working detection rules. In practice, most automated attempts at this only extract the simplest clues from the report --- bad IP addresses, domain names, and file hashes --- and turn them into block lists. This is a weak strategy, because attackers can change these simple clues within hours or days, so the resulting detections stop working almost as soon as they are deployed. Security teams describe this idea with the Pyramid of Pain: clues such as IP addresses and file hashes sit at the bottom of the pyramid and are cheap for an attacker to change, while clues about an attacker's behaviour and tooling --- their Tactics, Techniques and Procedures (TTPs) --- sit at the top and are expensive to change.

In this work we study whether feeding a report into a knowledge-graph retrieval system, Microsoft GraphRAG, rather than a standard vector-similarity retrieval system (Naive RAG), produces detection plans that rely more on these durable, top-of-pyramid clues. Both systems are given the same report, the same generation instructions, and the same language model to write the final plan; only the retrieval step differs. A second, cybersecurity-tuned language model then grades every plan against a ten-point rubric split across two tiers: foundation quality and Pyramid-of-Pain resilience.

In a detailed case study of one APT28 report, the GraphRAG plan kept firing at 100\% of its detections after every IP address, domain, and file hash in the report was rotated, while the Naive RAG plan kept firing at only 29\%. Repeating the comparison across nine real CTI reports from four vendors confirms the same pattern: GraphRAG plans consistently reach higher, harder-to-evade levels of the pyramid, even when the two systems end up close on total score. The results support treating knowledge-graph-aware retrieval as the architecturally correct foundation for automatically generating SOC-deployable hunting plans, while showing that the wording of the generation prompt matters almost as much as the retrieval back-end itself.
\end{abstract}


\maketitle

\section{Introduction}

\subsection{The problem: reports are hard to turn into detections}
When a security company such as CrowdStrike, Mandiant, EclecticIQ or Cyble publishes a report about a new attack campaign, the report usually contains everything a defender needs: the attacker's tools, their step-by-step methods, and concrete clues such as IP addresses or file hashes. In theory, a SOC could read the report and start defending against the same attack within hours. In practice this almost never happens automatically. A senior detection engineer has to read the whole report, pull out the useful clues, match them to known attack techniques, write detection queries for a tool such as Splunk or Microsoft Sentinel, and then go back and forth with the SOC to tune out false alarms. For a single high-quality vendor report this cycle typically takes two to ten engineer-days, so most newly published reports are never turned into a running detection at all.

Some teams try to automate this step with a language model that reads the report and writes a plan. A common approach is Retrieval-Augmented Generation (RAG): the model searches the report text for the passages most similar to the question being asked, and writes its answer from those passages~\cite{Lewis2020RAG}. The plans this produces look complete --- a campaign summary, a list of clues, ATT\&CK tables, detection paragraphs. Look closer, though, and most of the actual detection logic is anchored on the simplest, most literal clues in the report: file hashes, IP addresses, and domains. These are precisely the clues an attacker can change fastest, often inside 48 hours of the report going public, so the detections expire before the change-management ticket to deploy them even clears.

\subsection{Why some clues matter more than others}
Security researcher David Bianco's Pyramid of Pain~\cite{Bianco2014Pyramid} explains why this matters. It ranks detection clues by how much trouble they cause an attacker once defenders start blocking them. File hashes and IP addresses sit at the bottom: an attacker dodges a file-hash detection by recompiling the binary, which takes under an hour, and dodges an IP-based detection by renting a new server, which takes a day or two; either move costs next to nothing. Near the top of the pyramid sit Tactics, Techniques and Procedures (TTPs) --- the actual behavioural and tooling patterns of an attack, such as how a malicious document launches other programs on a victim machine. To dodge a detection built on TTPs, an attacker has to redesign how the whole operation works, which can take months. A hunting plan whose detections are dominated by the bottom three pyramid levels therefore has a useful life measured in hours; a plan whose detections sit at the top four levels keeps firing long after the report's specific indicators have gone stale.

\subsection{What this project asks}
This project compares two ways of automatically turning a CTI report into a threat hunting plan. Both use the same language model and the same generation instructions; the only difference is how each one searches the report before writing the plan. The first, Naive RAG, simply retrieves the report passages that look most similar in embedding space to the question being asked. The second, Microsoft GraphRAG~\cite{Edge2025GraphRAG}, first builds a small knowledge graph out of the report --- the threat actor, their malware, their infrastructure, and the relationships between these entities --- and then retrieves from that graph instead of from raw text.

The central question is simple: does searching a knowledge graph instead of searching text produce a plan that leans more on the durable, hard-to-evade clues near the top of the Pyramid of Pain? Three narrower research questions follow. \textbf{RQ1:} does GraphRAG retrieval push the dominant Pyramid level of a generated plan upward (L4--L7) compared with Naive RAG? \textbf{RQ2:} does GraphRAG retrieval increase the fraction of detections that keep firing after the adversary rotates every IOC disclosed in the report? \textbf{RQ3:} how sensitive is the GraphRAG-versus-Naive-RAG comparison to the generation prompt --- can a stricter prompt close the gap, or does the retrieval architecture dominate regardless?

\subsection{What this paper contributes}
This paper makes four contributions. First, a reproducible, fully on-premise pipeline that converts CTI PDFs into threat hunting plans through three retrieval back-ends --- GraphRAG Local Search, GraphRAG Global Search, and Naive vector RAG --- all driven by the same generation prompt and the same locally-hosted language model. Second, a multi-turn LLM-as-Judge harness built specifically for detection-engineering output, using a cybersecurity-domain judge model with explicit score floors, ceilings, and a step-by-step counting procedure designed to resist reward-hacking by well-written but indicator-thin plans. Third, two empirical evaluations: a single-report deep-dive that produces criterion-level scores for all three pipelines on an APT28 advisory, and a breadth experiment across nine real CTI reports drawn from four vendors. Fourth, a characterisation of two failure modes --- silent failure of GraphRAG Global Search on short reports, and JSON parsing failures in the small judge model on long output --- that are intrinsic to the pipeline rather than artefacts of the GraphRAG-versus-Naive-RAG comparison itself.

\section{Related Work}

Automating CTI operationalisation has attracted growing attention across four overlapping threads: extracting TTPs and techniques from unstructured reports, applying retrieval-augmented generation to CTI analysis and rule generation, building knowledge-graph-based reasoning systems over threat intelligence, and benchmarking how well LLMs perform SOC and threat-hunting tasks end to end. This project sits mainly at the intersection of the second and third threads but evaluates a different outcome variable --- the Pyramid-of-Pain durability of a generated hunting plan --- than most prior work, which reports extraction accuracy, rule-compilation rate, or task-completion reward instead.

\textbf{TTP extraction.} A large body of work treats CTI operationalisation as a classification problem: given a report, label which MITRE ATT\&CK~\cite{Mitre2023Attack} techniques it describes. TTPMapper~\cite{Ali2024TTPMapper} pairs two CyBERT classifiers, one trained on keyword-specific sentences and one on simplified and elaborated sentences, with a GPT-4o fallback for low-confidence cases, and reports 94.08\% accuracy across 202 ATT\&CK techniques --- the widest technique coverage among the systems it compares against. Kim et al.'s multi-step pipeline~\cite{Kim2025MultiStepTTP} instead splits the task into an LLM-based procedure-level \emph{Extractor}, an embedding-driven \emph{Technique Candidate Generator}, and an LLM \emph{Validator} that re-ranks candidates to suppress false positives, reaching an F1-score of 82.28\%. Büchel et al.'s USENIX Security 2025 systematisation of knowledge~\cite{Buechel2025SoK} re-implements a wide range of prior NLP approaches --- from named-entity recognition through generative LLMs --- in one shared evaluation setting; their central finding, that traditional NLP approaches can outperform embedder-based and generative approaches under realistic conditions and that existing approaches share a common performance ceiling, is a useful caution against assuming a newer architecture is automatically better. Sauerwein and Pfohl~\cite{Sauerwein2022TTPClassification} take an earlier NLP-plus-ML approach to the same problem, and Alam et al.'s LADDER~\cite{Alam2022LADDER} extracts attack patterns from CTI text and maps them to ATT\&CK phases for Android and enterprise campaigns. None of these five systems produces a deployable artefact such as a Splunk or Sentinel query; they stop at the technique label, closer here to an intermediate signal than a final output.

\textbf{RAG for CTI analysis and rule generation.} CT-RAG~\cite{Chandrakala2026CTRAG} and CyberLLM-FINDS~\cite{Iyer2026CyberLLMFinds} both extend RAG for CTI analysis rather than for hunting-plan generation. CT-RAG combines hybrid threat classier, risk prediction, retrieval-augmented summarized, and rule-augmented severity engine, to generate contextual representations of analysts workflows. The hybrid threat classifier fuses transformer, CNNs, and BiLSTM to derive a high accuracy attack classification, and risk prediction is performed through a DNN that scores context-aware risk from fused IoC and ATT\&CK-tactic embeddings. The rule-based severity engine improvises the severity score based on expert designed TTP rules. The evaluation on CTI-HAL dataset showed CTI-RAG outperformed that standalone methods. CyberLLM-FINDS fine-tunes Gemma-2B on synthetic cybersecurity instructions and layers a STIX-aware RAG-plus-graph module on top to improve multi-hop ATT\&CK-technique alignment, and runs an LLM-judged comparison of a Pure-RAG, a Graph+LLM and a GraphRAG+GNN configuration on a small set of MITRE queries. Three further systems push RAG closer to a deployable rule. LLMCloudHunter~\cite{Schwartz2024LLMCloudHunter} extracts detection-rule candidates from cloud CTI and compiles 99.18\% of them into valid Splunk queries; FALCON~\cite{Mitra2025Falcon} adds a self-reflection loop that produces Snort/YARA rules validated by a semantic scorer, reaching a mean analyst-rated relevance of 0.72 with 84\% inter-rater agreement; and ThreatPilot~\cite{Xu2024ThreatPilot} uses multi-hop GraphRAG-style reasoning to extract layered attack intelligence and auto-generate Sigma rules, reporting a $1.34\times$ F1 improvement over AttacKG on technique identification and raising attack-command execution rate from 50.3\% to 99.3\% when the extracted intelligence is used. CTI-REALM~\cite{Chakraborty2026CTIREALM} instead evaluates 16 frontier LLM agents on constructing detection rules against emulated attacks in live Linux, cloud and Kubernetes environments rather than against a static rubric. All four systems measure rule validity, compilation rate, or emulated-attack reward; none measures whether the resulting rule survives an adversary rotating the IOCs it was built from. The closest empirical precedent to this project's central comparison is Hamzic et al.~\cite{Hamzic2026BeyondRAG}, who evaluate four RAG architectures --- vector, graph-based, agentic query-correcting, and hybrid graph-text --- on 3{,}300 CTI question-answer pairs and find that their hybrid graph-text architecture improves performance by up to 35\% over vector-only RAG on multi-hop questions, with graph grounding generally helping structured factual queries. That result is the clearest existing evidence that graph-augmented retrieval helps CTI reasoning; this project asks the complementary question of whether the same architectural choice changes the durability of a generated detection artefact, not just the accuracy of a question-answering response.

\textbf{Knowledge graphs and GraphRAG for CTI.} CTI-Thinker~\cite{Yang2026CTIThinker} and CTIGen~\cite{Jin2025CTIGen} are the two systems closest in spirit to this project's use of graph structure. CTI-Thinker builds a CTI knowledge graph using in-context learning and LoRA-fine-tuned entity/relation extraction, then layers a GraphRAG-based reasoning engine on top for attack-intent inference and question answering; it reports higher precision, robustness and generalisability than prior extraction baselines, but --- like CT-RAG --- it targets knowledge-graph construction and reasoning quality rather than the durability of a downstream detection plan. CTIGen generates full malware-analysis CTI reports directly from decompiled binaries by combining static and dynamic analysis with a graph-based ATT\&CK grounding module, reporting 77.23\% ATT\&CK technique-identification accuracy and the discovery of 121 malicious functions not documented in the corresponding human-written reports; its graph is built from decompiled code rather than from a published vendor advisory. The graph-construction step both systems rely on has its own literature: AttacKG~\cite{Li2022AttacKG} was the first to extract technique knowledge graphs at scale, identifying 28{,}262 ATT\&CK techniques across 1{,}515 real-world reports; AttacKG+~\cite{Zhang2024AttacKGPlus} adds an LLM-based rewrite/parse/identify/summarise pipeline to upgrade these graphs with behavioural and temporal TTP labels; and CTINexus~\cite{Cheng2024CTINexus} uses in-context learning with hierarchical entity alignment to build CTI knowledge graphs from 150 reports without heavy fine-tuning. CyKG-RAG~\cite{Kurniawan2024CyKGRAG} and its successor AgCyRAG~\cite{Kurniawan2025AgCyRAG} integrate a structured cybersecurity knowledge base (CVE, CWE, CAPEC, ATT\&CK) with agentic vector-and-graph retrieval for security QA, and Han et al.~\cite{Han2025GraphRAGSurvey} frame GraphRAG as a query-processor/retriever/organiser/generator pipeline over graph-structured memory. None proceeds from a constructed knowledge graph to a behavioural hunting plan scored against IOC rotation.

\textbf{SOC-LLM benchmarks and domain-specialised models.} A parallel thread benchmarks how well general-purpose LLMs perform SOC work end to end. Habibzadeh et al.~\cite{Habibzadeh2025SOCSurvey} survey LLM use across the SOC lifecycle and flag multi-step, dynamic-decision-making reasoning as a recurring weakness. Two recent benchmarks quantify it directly: the Cyber Defense Benchmark~\cite{Chona2026CyberDefenseBenchmark} has LLM agents hunt for malicious events across 106 real attack procedures in Windows event-log corpora and finds the best frontier model identifies only 3.8\% of malicious events on average; and CyberTeam~\cite{Meng2025CyberTeam} shows that decomposing threat hunting into 30 standardised tasks across 9 operational modules outperforms open-ended agent reasoning. Bertiger et al.~\cite{Bertiger2025EvalDetectionRules} propose a holdout-set methodology for comparing LLM-generated detection rules against human-written ones, without a Pyramid-of-Pain-style durability axis. On the model side, SecureBERT~\cite{Aghaei2022SecureBERT}, SecureBERT~2.0~\cite{Aghaei2025SecureBERT2}, and the Foundation-Sec-8B technical report~\cite{Kassianik2025FoundationSecReport} (the base model behind the judge used in this work) show that cybersecurity-corpus pretraining materially improves security-text understanding, with Foundation-Sec-8B matching Llama-3.1-70B and GPT-4o-mini on several cybersecurity tasks despite its much smaller size.

\textbf{Positioning.} Across the systems surveyed above, the outcome measured is extraction accuracy, rule-compilation or validity rate, question-answering accuracy, knowledge-graph quality, or emulated-attack task reward. Hamzic et al.~\cite{Hamzic2026BeyondRAG} come closest to this project's architectural question by showing graph retrieval beats vector RAG on CTI question answering, and ThreatPilot~\cite{Xu2024ThreatPilot} comes closest to its output format by generating rules with GraphRAG-style reasoning; neither asks whether the resulting artefact survives an adversary rotating their IOCs, which is the Pyramid-of-Pain-native way a SOC actually judges a hunting plan's operational lifespan. This project isolates that variable by holding the generation model, the generation prompt, and the judge model constant and varying only the retrieval back-end (GraphRAG Local/Global versus Naive RAG), then scoring the result on a rubric built directly around Pyramid-of-Pain durability rather than technique-label accuracy, rule validity, or QA accuracy.

\section{Background}

\subsection{The CTI Lifecycle and Threat Hunting}
CTI as practised in modern SOCs follows a five-stage lifecycle: collection (vendor feeds, ISACs, pastebin scrapers), processing (deduplication, enrichment, attribution), analysis (campaign clustering, ATT\&CK tagging), dissemination (intelligence reports, IOC feeds in structured formats such as STIX~\cite{Oasis2021Stix}, hunt packages), and feedback (telemetry validation, false-positive review); a recent survey catalogues LLM applications across this lifecycle, including log analysis, triage and detection support~\cite{Habibzadeh2025SOCSurvey}. The dissemination artefact most commonly consumed by detection engineers is the long-form vendor report. The hunting plans evaluated in this work sit at the dissemination-to-feedback bridge: an SOC L2 analyst should be able to take the plan, paste each detection into Splunk or Microsoft Sentinel, and either fire on real telemetry or be tuned out within 72 hours. A threat hunting plan therefore needs three properties that a generic CTI summary does not: (i) every detection must reference a real telemetry source and field name; (ii) every detection must declare the Pyramid level it targets, so the SOC can prioritise; (iii) every behavioural detection must declare a false-positive baseline, at minimum a named exclusion list or a numeric threshold.

\subsection{Retrieval-Augmented Generation (Naive RAG)}
Retrieval-Augmented Generation~\cite{Lewis2020RAG} augments a generative LLM with documents retrieved from an external corpus at query time. The Naive variant used as the comparison baseline here splits the source document into 600-word chunks with 80-word overlap, embeds each with Qwen3-Embedding-8B~\cite{Qwen2025Embedding} into a LanceDB vector index for approximate-nearest-neighbour search~\cite{Johnson2021Faiss}, and at query time concatenates the top-8 chunks by cosine similarity into the LLM context. This works well when the answer is contained in a small contiguous span of source text; it works poorly when the answer requires reasoning over a graph of entities --- for example, which of the C2 domains in a report shares infrastructure with the spear-phishing infrastructure --- because the relevant entities may never co-occur in any single retrieved chunk. This failure mode mirrors the multi-hop question-answering setting studied outside the security domain~\cite{Yang2018HotpotQA}, where the answer requires combining evidence spread across passages that individually look unrelated to the question, and it matches the CTI-specific finding that vector-only RAG degrades sharply on multi-hop CTI questions relative to graph-grounded retrieval~\cite{Hamzic2026BeyondRAG}.

\subsection{GraphRAG: Local and Global Search}
Microsoft GraphRAG~\cite{Edge2025GraphRAG}, and graph-augmented retrieval more generally~\cite{Han2025GraphRAGSurvey}, augment standard RAG with an explicit knowledge graph constructed at indexing time. The indexing pipeline runs five stages: (i) text-unit chunking, (ii) LLM-driven entity and relationship extraction with a domain-tunable entity-type list, (iii) summarisation of each entity's mentions across the corpus, (iv) Leiden community detection~\cite{Traag2019Leiden} over the entity-relationship graph at multiple resolution levels, and (v) LLM-driven summarisation of each community at every level. The output is a heterogeneous structure of entities, relationships, community summaries and the original text units, stored as Parquet files plus a LanceDB vector store.

GraphRAG exposes two retrieval strategies relevant to this work. \textbf{Local Search} starts from the user query, identifies the most semantically relevant entities by embedding similarity, then walks the entity-relationship graph to gather the relationships among those entities, the text units that mention them, and the community summaries that contain them; it is a natural fit for threat-hunting questions anchored on a campaign's named entities (the actor, the implant, the C2 infrastructure). \textbf{Global Search} ignores entity-level retrieval and instead synthesises an answer by querying the LLM with batches of community summaries at a chosen community level, then combines the partial answers in a second LLM pass; it excels at corpus-wide sense-making but is sensitive to community structure, and returns a near-empty answer when a source document produces fewer than three or four communities.

\subsection{The Pyramid of Pain Framework}
The Pyramid of Pain~\cite{Bianco2014Pyramid} orders detection artefacts by the cost an adversary must pay to evade them, across seven levels: L1 file hashes (adversary cost to rotate: under an hour, recompile or repack), L2 IP addresses (24--48 hours, rotate VPS/Tor/proxy), L3 domain names (minutes, re-register or fast-flux), L4 network artefacts (days, redesign the C2 protocol), L5 host artefacts (days--weeks, rewrite the implant), L6 tool fingerprints (weeks, re-tool), and L7 TTPs (months, redesign tradecraft). The judge rubric used in this work encodes the framework's central claim --- that detection quality is not a scalar but determined by where a detection sits on this seven-level pyramid --- through a Tier~2 score dominated by Pyramid-level metrics and a tie-breaker that favours the plan with the higher Tier~2 score even when its Tier~1 score is lower.

\subsection{LLM-as-Judge for Detection-Engineering Outputs}
LLM-as-Judge~\cite{Zheng2023Judge} uses a separate LLM as the evaluator of generated text against a rubric. The technique is well established for general chat-quality evaluation but underexplored for detection-engineering outputs, which are harder to judge in three ways: the criteria are technical (the difference between an L2 and an L7 detection is unambiguous to a domain expert but invisible to a general-purpose judge); the criteria interact (specificity reinforces detection-readiness, while intelligence-grounding gates everything else); and the output is long and domain-specific enough that an 8B-parameter judge running locally is at the edge of its capability, with observed failure modes including arithmetic errors when summing tier totals and JSON malformation on long responses. Both failure modes are addressed in this work: the former by programmatic recomputation of tier totals from the individual criterion scores, and the latter by splitting the evaluation into four short turns.

\section{System Design}

\subsection{End-to-End Architecture}
\begin{figure*}[t]
  \centering
  \includegraphics[width=0.6\textwidth]{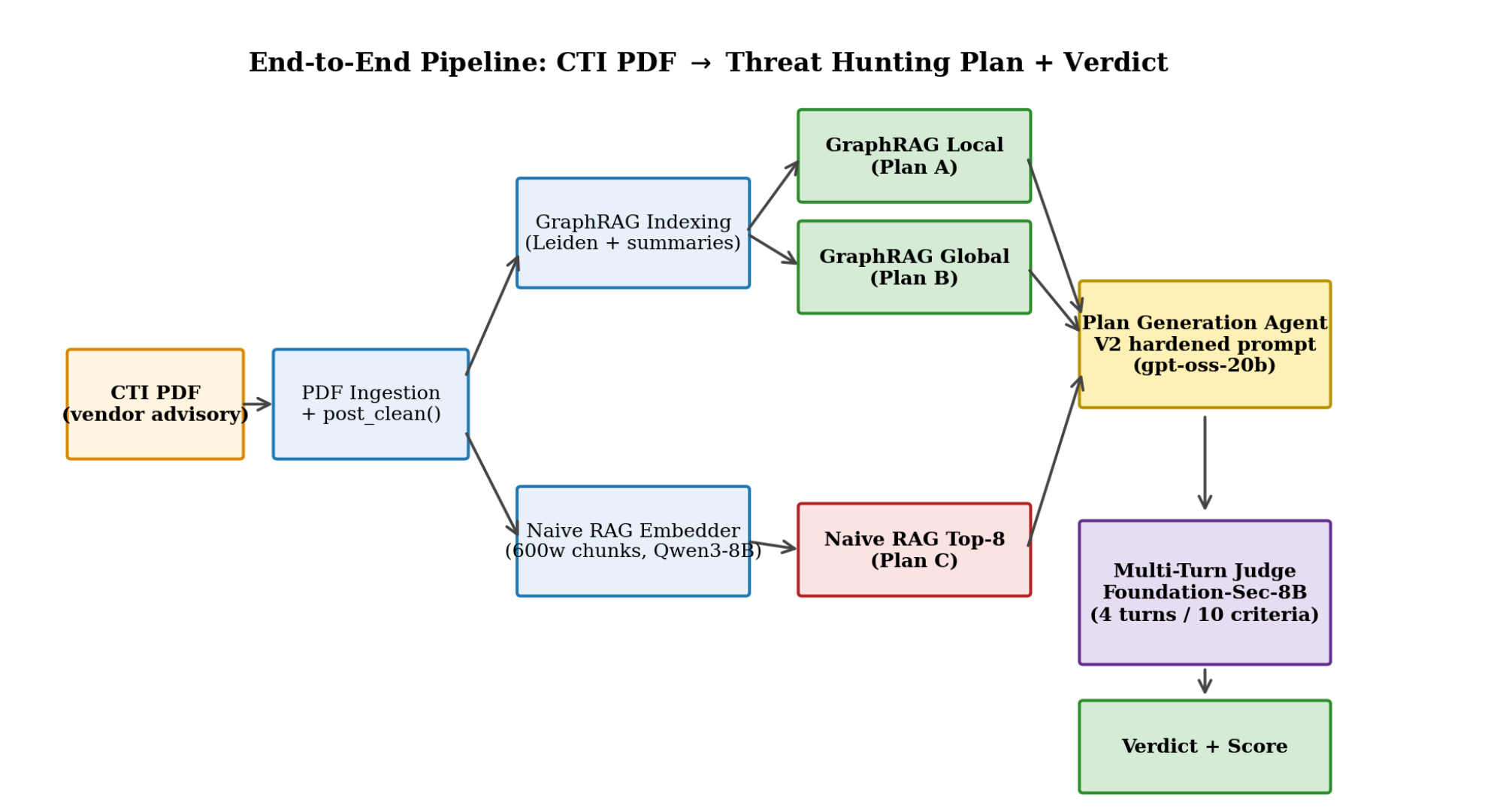}
  \caption{End-to-end pipeline. The same cleaned text is consumed by GraphRAG indexing and the Naive RAG embedder. All three plans are generated with the same prompt and the same gpt-oss-20b completion model, then judged by Foundation-Sec-8B-Instruct in four sequential turns.}
  \Description{A flow diagram showing a CTI PDF going through PDF ingestion and cleanup, then splitting into a GraphRAG indexing path producing GraphRAG Local (Plan A) and GraphRAG Global (Plan B), and a Naive RAG embedder path producing Naive RAG Top-8 (Plan C). All three plans feed into a Plan Generation Agent using the V2 hardened prompt and gpt-oss-20b, which feeds a Multi-Turn Judge using Foundation-Sec-8B across four turns and ten criteria, producing a final Verdict and Score.}
  \label{fig:pipeline}
\end{figure*}

The pipeline takes a CTI PDF as input and emits three threat hunting plans (one per retrieval back-end), four judge transcripts, and a verdict. Figure~\ref{fig:pipeline} shows the end-to-end flow. The pipeline runs entirely on a single workstation with no external API calls. Three vLLM-served local endpoints provide completion (gpt-oss-20b~\cite{OpenAI2025GptOss}), embedding (Qwen3-Embedding-8B~\cite{Qwen2025Embedding}) and judging (Foundation-Sec-8B-Instruct~\cite{Cisco2024FoundationSec}); Table~\ref{tab:endpoints} summarises the model assignments.

\begin{table}[t]
  \caption{Model and endpoint configuration. All three are served by vLLM on a single GPU host.}
  \label{tab:endpoints}
  \small
  \begin{tabular}{@{}p{1.5cm}p{2.3cm}p{1.6cm}@{}}
    \toprule
    Component & Model & Endpoint \\
    \midrule
    Plan generation & openai/gpt-oss-20b & localhost:8001 \\
    Embedding & Qwen/Qwen3-Embedding-8B & localhost:8002 \\
    Judge & fdtn-ai/Foundation-Sec-8B-Instruct & localhost:8000 \\
    \bottomrule
  \end{tabular}
\end{table}

\subsection{PDF Ingestion and Cleanup}
Vendor CTI PDFs are typically authored in InDesign or Word and exported with watermarks, repeated headers and TLP markers that pdfminer extracts as isolated text fragments. These fragments severely degrade GraphRAG entity extraction because the LLM treats each one-character watermark line as a candidate entity; the original extractor produced unusable graphs for two of the nine input reports (Cyble OpSindoor and the Vishing/Help-desk MSC). A three-pass post-cleaning routine, \texttt{\_post\_clean()}, fixes this. Pass one drops lines matching \texttt{\^{}\textbackslash s*[A-Za-z]\{1,2\}\textbackslash s*\$}, removing the single-character fragments produced by vertically rendered watermarks. Pass two drops anchored chrome patterns: lines starting with \texttt{TLP:}, \texttt{Copyright \textbackslash d\{4\}}, \texttt{Page N}, and \texttt{DD/MM/YYYY Page N}. Pass three drops any non-empty line appearing three or more times in documents longer than three pages, or twice or more in shorter documents, capturing repeated footers and running headers without dropping legitimate repeated body content.

\subsection{GraphRAG Indexing}
Indexing uses the Microsoft GraphRAG reference implementation~\cite{Edge2025GraphRAG} with a CTI-tuned configuration. Entity extraction is invoked with eight domain entity types: actor, malware, technique, vulnerability, domain, ip, file, sector. The relationship extractor is run twice --- first to extract direct mentions, then with a gleanings pass that probes the LLM for relationships missed by the first pass. Community detection runs at three Leiden resolution levels (0, 1, 2) so that subsequent retrieval can choose the granularity. The output is six Parquet files (entities, relationships, text-units, communities, community-reports, documents) plus a LanceDB vector store of the entity description embeddings. Indexing parameters: chunk size 600 words, chunk overlap 80 words, top-$K$ retrieval 8, GraphRAG community-summary level 2. These values were tuned once during early pipeline development and are held fixed for all experiments here.

\subsection{Three Retrieval Back-ends}
Table~\ref{tab:backends} summarises the three retrieval back-ends evaluated, all driving the same generation prompt and the same generation model. As determined in the project scope, Plans A and B are reported as a single GraphRAG family --- ``GraphRAG'' in win-count tallies refers to the better-scoring of the two on each report. Naive RAG (Plan C) is the comparison baseline.

\begin{table*}[t]
  \caption{The three retrieval back-ends evaluated in this work. All three drive the same generation prompt and the same generation model.}
  \label{tab:backends}
  \begin{tabular}{@{}p{3.1cm}p{6.5cm}p{6.7cm}@{}}
    \toprule
    Pipeline & Retrieval mechanism & Context construction \\
    \midrule
    Plan A (GraphRAG Local) & Entity-relationship graph traversal seeded by query embedding & Selected entities + their relationships + text units mentioning them + their parent community summaries \\
    Plan B (GraphRAG Global) & Map-reduce over level-2 community summaries & Per-community partial answers concatenated and re-summarised by the LLM \\
    Plan C (Naive RAG) & Top-8 cosine similarity over 600-word chunks & Top-8 chunks concatenated verbatim before the prompt \\
    \bottomrule
  \end{tabular}
\end{table*}

\subsection{Plan-Generation Agent}
The plan-generation agent is the same prompt for all three pipelines. The full text of the V2 hardened prompt's non-negotiable rules is reproduced in Appendix~\ref{app:prompt}; in summary it instructs the model to act as a principal threat intelligence analyst and produce seven sections: a campaign summary, a behavioural detection chain section labelled \textsc{Primary}, an IOC hunting list explicitly labelled \textsc{Fragile}, an ATT\&CK coverage table, an IOC rotation resilience analysis, a priority hunting actions table, and a false-positive mitigation section. Every detection in section two must carry a Pyramid-level tag and a field-level Splunk SPL query plus a Microsoft Sentinel KQL query.

\subsection{Multi-Turn LLM-as-Judge}
The judge runs as four separate API calls to keep each turn under approximately 14{,}000 tokens --- a hard ceiling for the 8B Foundation-Sec model~\cite{Kassianik2025FoundationSecReport} running with a 16K context window. Turns 1, 2 and 3 each evaluate one plan against the ten-criterion rubric and emit a JSON object containing per-criterion scores, per-criterion reasoning, strengths and weaknesses. Turn 4 receives only the structured JSON outputs from the first three turns and synthesises a verdict identifying the winning plan with explicit reference to the Tier~2 (Pyramid resilience) tie-breaker. Tier totals returned by the model are programmatically discarded and recomputed from the individual scores; the 8B model was repeatedly observed to compute $9 \times 10 = 90$ instead of summing the actual scores, and the recomputation eliminates this entire failure class.

\section{Experimental Methodology}

\subsection{CTI Report Corpus}
The corpus consists of nine real CTI reports drawn from four vendors (CrowdStrike, Cyble, EclecticIQ, Mandiant) plus one open-source PDF advisory. The reports cover a range of campaign types --- nation-state APT, e-crime, vulnerability advisory, social engineering --- and a range of lengths from approximately two pages (LAPSUS\$ insider recruitment) to approximately twenty-five pages (LABYRINTH CHOLLIMA TxRLoader). The deep-dive single-report experiment uses the APT28 LayeredMesh advisory~\cite{CrowdStrike2026Fancybear}. Table~\ref{tab:corpus} lists all nine reports.

\begin{table}[t]
  \caption{The nine CTI reports used in the breadth experiment.}
  \label{tab:corpus}
  \small
  \begin{tabular}{@{}p{4.1cm}p{1.5cm}p{1.9cm}@{}}
    \toprule
    Report (short name) & Vendor & Campaign type \\
    \midrule
    FancyBear / APT28 LayeredMesh & CrowdStrike & Nation-state APT (Russia) \\
    LABYRINTH CHOLLIMA TxRLoader~\cite{CrowdStrike2025Txrloader} & CrowdStrike & Nation-state APT (DPRK), crypto \\
    LABYRINTH CHOLLIMA macOS & CrowdStrike & Nation-state APT (DPRK), macOS \\
    Cyble OpSindoor (APT36) & Cyble & Nation-state APT (Pakistan) \\
    Vishing / Help-desk MSC & EclecticIQ & Social engineering / e-crime \\
    Zimbra LFI advisory~\cite{CrowdStrike2026Zimbra} & CrowdStrike & Vulnerability / mass exploitation \\
    Renegade Jackal / Micropsia & EclecticIQ & Nation-state APT (Middle East) \\
    LAPSUS\$ insider recruitment & EclecticIQ & E-crime / insider threat \\
    BADBOX2 Android backdoor & EclecticIQ & Supply-chain / mobile malware \\
    \bottomrule
  \end{tabular}
\end{table}

\subsection{Generation Prompt: V1 Baseline \texorpdfstring{$\rightarrow$}{to} V2 Hardened}
Two generation prompts were evaluated. The V1 baseline is a generic seven-section threat-hunting prompt; the V2 hardened prompt adds five non-negotiable rules --- source fidelity, query executability, Pyramid tagging, ATT\&CK depth, and behavioural priority --- that materially alter generation behaviour (full text in Appendix~\ref{app:prompt}). The V1$\rightarrow$V2 difference is the largest single driver of generation quality observed in this work: the V1 baseline produced plans that scored 6, 24 and 23 out of 100 for Plans A, B and C on the APT28 deep-dive; the V2 hardened prompt lifts the same plans to 80, 79 and 78 --- a fifty-five to seventy-four point swing on identical retrieval back-ends. Section~\ref{sec:v1v2} discusses this calibration effect in detail.

\subsection{Judge Rubric and Score Floors / Ceilings}
Each plan is scored on ten criteria across two tiers. Tier 1 captures foundation quality (60 points); Tier 2 captures Pyramid-of-Pain resilience (40 points). Table~\ref{tab:rubric} summarises the criteria and their key floor or ceiling.

\begin{table*}[t]
  \caption{The ten-criterion judge rubric. Floors enforce minimum credit when concrete evidence exists; ceilings cap scores when discriminating evidence is absent.}
  \label{tab:rubric}
  \small
  \begin{tabular}{@{}p{0.5cm}p{3.4cm}p{6.3cm}p{6.6cm}@{}}
    \toprule
    Tier & Criterion (max 10) & What it measures & Key floor / ceiling \\
    \midrule
    1 & completeness & Kill-chain phase coverage with actual queries & $-2$ per missing phase \\
    1 & specificity & Campaign-specific observables vs.\ generic TTPs & $\geq 3$ if any IOC cited; $\leq 3$ if none \\
    1 & attck\_coverage & Fraction of ATT\&CK IDs paired with detection logic & $\leq 4$ if IDs listed without queries \\
    1 & detection\_readiness & Copy-paste readiness for Splunk / Sentinel / Elastic & $\geq 3$ if any field-level query; $\leq 4$ if all prose \\
    1 & fp\_mitigation & Named exclusions and numeric thresholds per detection & $\leq 2$ if only ``tune for your environment'' \\
    1 & intelligence\_grounding & Zero hallucinated IOCs / ATT\&CK IDs & $\leq 4$ if any unverifiable claim \\
    \midrule
    2 & pyramid\_level & Dominant Pyramid level across all detections & $\leq 4$ if $>$60\% are L1--L3 \\
    2 & detection\_durability & Count of detections surviving full IOC rotation & $\leq 3$ if $<$30\% survive \\
    2 & ttp\_behavioral\_depth & Sophistication of L4--L7 behavioural detections & $\leq 3$ if no behavioural detections \\
    2 & ioc\_resistance & Architectural resilience: ordering, fallback, labelling & $\leq 3$ if $<$30\% survive rotation \\
    \bottomrule
  \end{tabular}
\end{table*}

\paragraph{The 0--10 scale and its anchors.} Every criterion is scored on the same 0--10 integer scale, and tier and grand totals are arithmetic sums, granular enough to discriminate near-equivalent plans without the false precision a 0--100 scale would imply at an 8B judge's resolution. Each band carries an explicit anchor: 1--2 is absent or pure boilerplate (``hunt for lateral movement''); 3--4 is present but campaign-generic, naming a real technique with nothing from the report cited; 5 is adequate, with campaign-specific detail but notable gaps or vague queries; 6--7 is good, campaign-specific and mostly actionable bar one gap such as a missing FP note; 8 is strong --- specific, copy-paste ready, campaign-tied, would survive a basic red-team review; 9--10 is exceptional, requiring multiple specific examples, field-level queries, behavioural depth and explicit FP baselines, with 9 requiring at least three cited examples and 10 deliberately near-unreachable. The 0--2 floor is content a tier-1 SOC lead would reject on sight; the 9--10 ceiling is content that lead would deploy unmodified.

\paragraph{Why a two-tier split, and why 60/40.} The 60/40 split directly encodes the project hypothesis. A plan that is unreadable, ungrounded or dead-on-arrival is worthless regardless of Pyramid position, so Tier 1 sets the deployability floor and takes the larger share (60 of 100); Tier 2 then asks the question the project cares about --- does the plan keep firing after infrastructure rotation? --- and carries enough weight (40 of 100) that two plans differing only in Pyramid composition diverge by eight to twelve grand-total points, well outside single-judge-run noise. Tier 1's six criteria are the independent checks a senior SOC engineer runs on an incoming detection package: kill-chain coverage, specificity, ATT\&CK grounding, query executability, FP containment, and absence of hallucination. Tier 2's four criteria are deliberately non-orthogonal --- where detections sit, how many survive rotation, how sophisticated the durable subset is, and whether the plan is architecturally resilient overall --- so that a plan scoring high on all four is incontestably Pyramid-resilient even under judge noise.

\paragraph{Floors, ceilings, and the 30\%/60\% thresholds.} Five hard floors are re-applied programmatically: a campaign-specific IOC citation lifts specificity to $\geq 3$; a field-level query lifts detection\_readiness; an explicit FP baseline lifts fp\_mitigation; a behavioural detection lifts ioc\_resistance; an ATT\&CK ID with an attached observable lifts attck\_coverage. Pilot V1 runs on the APT28 advisory showed the collapse mode these floors prevent: the judge scored specificity 0/10 despite four campaign-specific IOCs being cited verbatim, because two prose-only queries elsewhere pulled the holistic impression down. Five hard ceilings are calibrated symmetrically: detection\_readiness and attck\_coverage cap at $\leq 4$ when all queries are prose or ATT\&CK IDs are decorative; intelligence\_grounding caps at $\leq 4$ when any IOC is unverifiable; ioc\_resistance and detection\_durability cap at $\leq 3$ when fewer than 30\% of detections survive rotation; fp\_mitigation caps at $\leq 2$ when the plan offers only ``tune for your environment,'' the canonical SOC red flag for an unengineered detection. The 60\% threshold on pyramid\_level reflects that a fragile-dominant plan is operationally fragile even with scattered durable detections; the 30\% threshold on detection\_durability and ioc\_resistance captures that a plan losing three-quarters of its surface in the first 48 hours is functionally dead at deployment. Both thresholds are pyramid-derived, not retrieval-derived, so they cannot bias the comparison toward either pipeline.

\paragraph{Counting procedure and recomputation.} The judge system prompt prescribes a six-step procedure executed before any score is written: list every detection and its Pyramid level; count L1--L3 versus L4--L7 detections; count how many would still fire after a complete rotation of campaign IPs, domains and hashes; apply the counts-conditioned ceilings; assign each criterion score and apply the floors; and write per-criterion reasoning in a fixed quote-then-justify-then-score format. This combination reduces criterion-score variance across re-runs by roughly a factor of two, the largest stability gain observed during rubric calibration. Tier totals and the grand total are then discarded and recomputed programmatically from the per-criterion scores by \texttt{\_evaluate\_single\_plan()}, eliminating the $9\times10=90$-style arithmetic error seen in raw judge output. This same recomputation produces the 16/100 figure that recurs throughout the breadth experiment as the GraphRAG Global silent-fail floor: when Global Search returns its 75-character error string, the floors still mechanically credit it with $\geq 3$ on specificity, attck\_coverage and ioc\_resistance whenever any single ATT\&CK ID and behavioural detection appear, which they typically do even in the empty fallback text. A score at or near 16 is therefore a silent pipeline failure, not a genuine quality reading.

\subsection{Win Determination: GraphRAG vs Naive RAG}
The headline comparison in this work is GraphRAG (the family) versus Naive RAG (Plan C). The GraphRAG result for a given report is the better-scoring of Plan A (Local Search) and Plan B (Global Search), reflecting how a real SOC would deploy the system: index the report once, run both retrieval strategies, and surface the higher-quality plan. Two design decisions matter here. First, judge-parsing failures, where the 8B model emits malformed JSON, are reported as failures and not as 0/100; earlier reporting collapsed both cases into 0/100 and overstated Naive RAG's win count. Second, GraphRAG Global silent failures are reported as a 16/100 floor rather than a genuine score, since that is what the rubric's floors mechanically produce on a near-empty plan.

\section{Results}

\subsection{V1 to V2 Calibration Effect}
\label{sec:v1v2}
\begin{figure*}[t]
  \centering
  \includegraphics[width=0.62\textwidth]{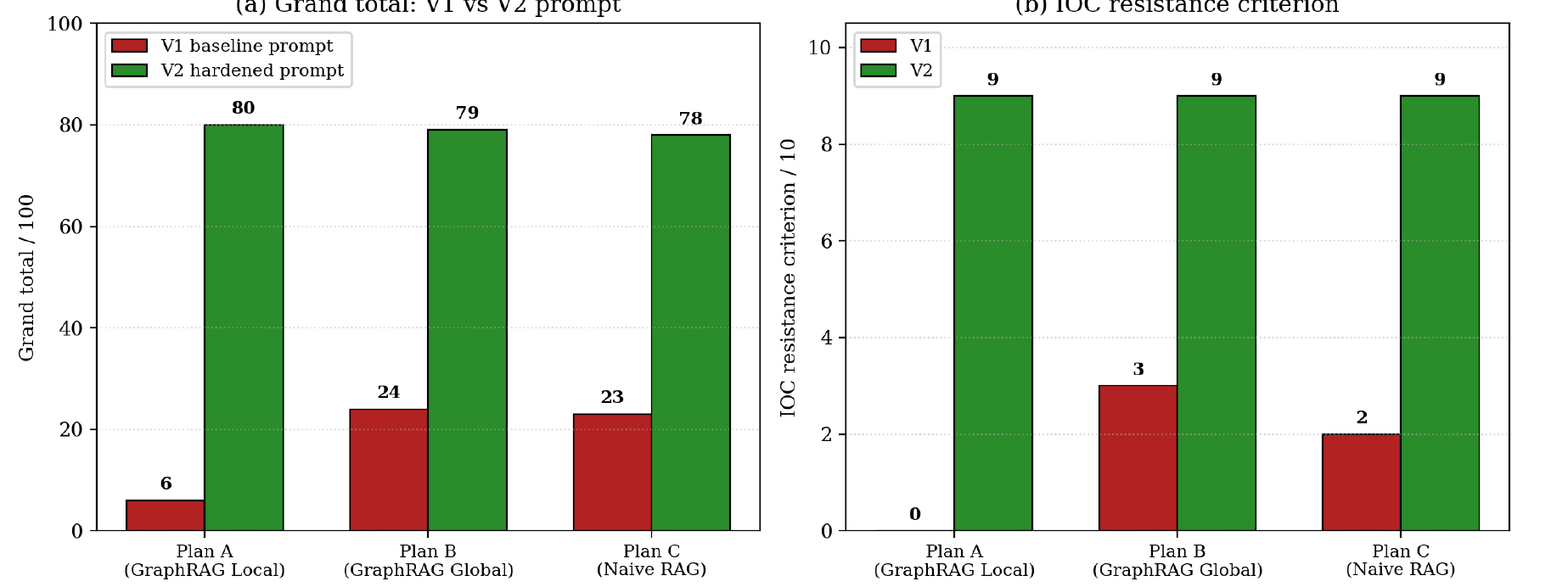}
  \caption{Effect of replacing the V1 generic prompt with the V2 hardened prompt on the APT28 deep-dive. All three pipelines move from sub-25/100 to mid-to-high 70s/100. The IOC-resistance criterion moves from 0/10 to 9/10.}
  \Description{Two bar charts. The left chart shows grand total scores out of 100 for Plans A, B and C under the V1 baseline prompt (6, 24, 23) versus the V2 hardened prompt (80, 79, 78). The right chart shows the IOC-resistance criterion out of 10 under V1 (0, 3, 2) versus V2 (9, 9, 9) for the same three plans.}
  \label{fig:v1v2}
\end{figure*}

Holding retrieval back-end, generation model and judge model fixed and changing only the generation prompt produces the change shown in Figure~\ref{fig:v1v2} and Table~\ref{tab:v1v2}.

\begin{table}[t]
  \caption{V1 to V2 calibration effect on the APT28 single-report deep-dive.}
  \label{tab:v1v2}
  \small
  \begin{tabular}{@{}p{3.5cm}p{1.1cm}p{1.1cm}p{1.1cm}@{}}
    \toprule
    Metric & V1 & V2 & Delta \\
    \midrule
    Plan A grand total / 100 & 6 & 80 & $+74$ \\
    Plan B grand total / 100 & 24 & 79 & $+55$ \\
    Plan C grand total / 100 & 23 & 78 & $+55$ \\
    IOC resistance (all plans) /10 & 0--3 & 9 & $+6$ to $+9$ \\
    Plan A dominant Pyramid level & L3 & L7-TTP & fragile$\rightarrow$durable \\
    Plan C dominant Pyramid level & L3 & L4 & fragile$\rightarrow$durable \\
    Plan A length (chars) & 9{,}198 & 13{,}441 & $+46\%$ \\
    Plan C length (chars) & 12{,}719 & 17{,}851 & $+40\%$ \\
    \bottomrule
  \end{tabular}
\end{table}

Three observations follow. First, the V1 prompt produced plans whose grand totals are within experimental noise of one another (24 vs 23 for Plans B and C); the prompt was so under-constrained that the retrieval back-end made almost no difference. Second, the V2 prompt produces grand totals that are likewise close (80, 79, 78) but the underlying Tier 2 composition diverges sharply (Section~\ref{sec:pyramid-dist}). Third, the V1-to-V2 move is much larger than any retrieval-back-end move, indicating that prompt engineering is at least as important as retrieval architecture for this class of generation task.

\subsection{APT28 Single-Report Deep-Dive}
The APT28 LayeredMesh advisory~\cite{CrowdStrike2026Fancybear} was selected for the deep-dive because it is dense enough to exercise GraphRAG community detection (eight pages, approximately three thousand entity mentions) and because it cleanly contains both behavioural primitives (Outlook macro persistence, Foreshadow side-channel exploitation, Cobalt Strike-derived implants) and traditional IOCs (six C2 IPs, four C2 domains, multiple file hashes). Table~\ref{tab:apt28} reproduces the full per-criterion V2 scores.

\begin{table}[t]
  \caption{Per-criterion scores for the APT28 V2 deep-dive. Tier 2, the Pyramid resilience tier, separates the plans more cleanly than Tier 1.}
  \label{tab:apt28}
  \small
  \begin{tabular}{@{}p{3.4cm}p{1.1cm}p{1.1cm}p{1.1cm}@{}}
    \toprule
    Criterion & A (Local) & B (Global) & C (Naive) \\
    \midrule
    completeness & 7 & 7 & 7 \\
    specificity & 8 & 8 & 8 \\
    attck\_coverage & 9 & 8 & 9 \\
    detection\_readiness & 7 & 7 & 7 \\
    fp\_mitigation & 6 & 9 & 6 \\
    intelligence\_grounding & 10 & 10 & 10 \\
    \midrule
    pyramid\_level & 9 & 6 & 7 \\
    detection\_durability & 8 & 8 & 8 \\
    ttp\_behavioral\_depth & 7 & 7 & 7 \\
    ioc\_resistance & 9 & 9 & 9 \\
    \midrule
    \textbf{Tier 1 total /60} & 47 & 49 & 47 \\
    \textbf{Tier 2 total /40} & 33 & 30 & 31 \\
    \textbf{Grand total /100} & \textbf{80} & 79 & 78 \\
    \bottomrule
  \end{tabular}
\end{table}

Plan A wins on the Tier 2 tie-breaker by reaching dominant L7-TTP. The grand-total margin is small (80 vs 78) but the architectural difference between an L7-dominant plan and an L4-dominant plan is operationally large: the L7 plan continues to fire on adversary tradecraft after a complete IOC rotation, while the L4 plan loses 71\% of its detections to the same rotation.

\subsection{Pyramid-of-Pain Distribution}
\label{sec:pyramid-dist}
\begin{figure*}[t]
  \centering
  \includegraphics[width=0.6\textwidth]{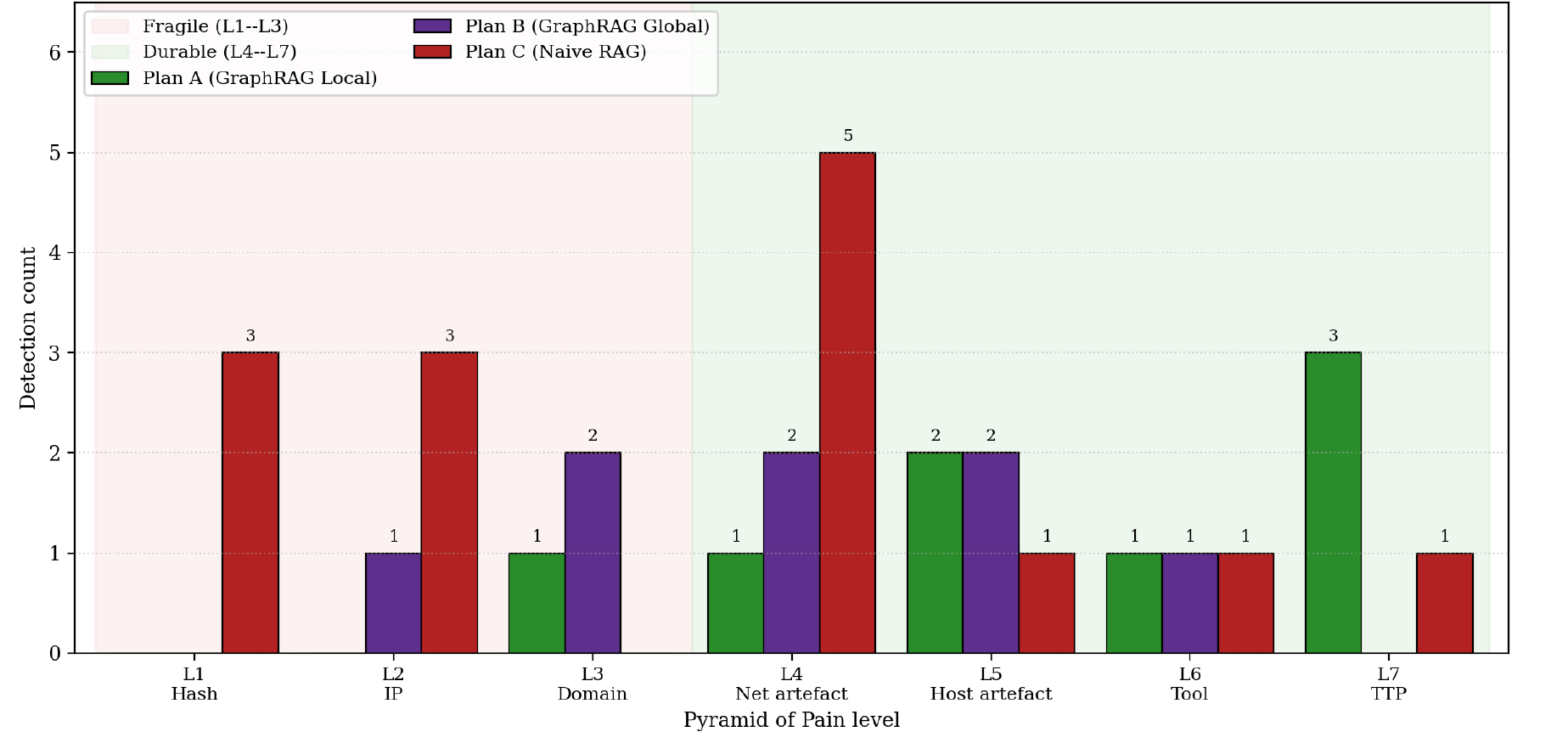}
  \caption{Aggregate Pyramid-of-Pain distribution across the V2 deep-dive plus the Run~2 batch. GraphRAG Local concentrates detections in L5--L7 (host artefacts, tools, TTPs); Naive RAG concentrates detections at L4 with a long fragile tail at L1--L3.}
  \Description{A grouped bar chart with seven Pyramid-of-Pain levels on the x-axis (L1 Hash through L7 TTP), shaded pink for the fragile L1-L3 region and green for the durable L4-L7 region. Plan C (Naive RAG, red bars) peaks at 5 detections at L4 and has counts of 3, 3, 0 at L1-L3. Plan A (GraphRAG Local, green bars) peaks at 3 detections at L7. Plan B (GraphRAG Global, purple bars) peaks at 2 detections spread across L3-L5.}
  \label{fig:pyramid}
\end{figure*}

Figure~\ref{fig:pyramid} shows the aggregate distribution of detections across the seven Pyramid levels for the three pipelines on the V2 deep-dive plus the Run~2 breadth batch, with the boundary between fragile (L1--L3) and durable (L4--L7) shaded. GraphRAG Local detections cluster at L5--L7 with a meaningful tail at L4. GraphRAG Global is heaviest at L3--L5 because community summaries surface IOC clusters as well as relationship clusters. Naive RAG is heaviest at L4, driven by the network-artefact chunks that dominate cosine-similarity retrieval, but retains a fragile L1--L3 tail of 45\% of its detections --- precisely the category adversary IOC rotation eliminates.

\subsection{IOC Rotation Survival}
Detection durability under IOC rotation is the headline operational metric of this work. For each plan, every detection was classified by whether it would still fire after a complete rotation of the campaign IOCs (all IPs, domains and file hashes substituted). Table~\ref{tab:rotation} reports the result on the APT28 V2 deep-dive.

\begin{table}[t]
  \caption{IOC rotation survival on the APT28 V2 deep-dive. GraphRAG plans retain 100\% of their detections after IOC rotation; Naive RAG retains 29\%.}
  \label{tab:rotation}
  \small
  \begin{tabular}{@{}p{2.6cm}p{1.05cm}p{1.05cm}p{0.85cm}p{1.4cm}@{}}
    \toprule
    Plan & Total & Surviving & Surv.\ \% & Dominant level \\
    \midrule
    A --- GraphRAG Local & 6 & 6 & 100\% & L7-TTP \\
    B --- GraphRAG Global & 8 & 8 & 100\% & L4--L6 \\
    C --- Naive RAG & 14 & 4 & 29\% & L4 \\
    \bottomrule
  \end{tabular}
\end{table}

This is the cleanest evidence in support of RQ2. The GraphRAG plans are not merely slightly more behavioural; they retain their entire detection surface after the adversary rotates infrastructure, while Naive RAG loses two-thirds of its surface to the same event. Note that the absolute count of GraphRAG detections is lower (6--8 vs 14); the comparison favouring GraphRAG is on durability, not volume.

\subsection{Breadth Experiment: All Nine CTI Reports}
To test how far the deep-dive findings generalise, the same pipeline was run over the full nine-report corpus in two batches. Run 1 used the original PDF extractor; Run 2 used the cleaned-up extractor with watermark and chrome-line removal. The two runs are reported together so that every report contributes to the comparison even when one run failed for it. Table~\ref{tab:breadth} reports the full grid.

\begin{table*}[t]
  \caption{Full breadth-experiment grid. Run 1 (R1) used the original PDF extractor; Run 2 (R2) used the cleaned extractor. ``---'' denotes a run that did not complete for that report. ``*'' marks a 0-score caused by a judge JSON parsing failure rather than a generation failure. ``Best GR'' and ``Best Naive'' are the maxima over the two runs and are used in the win count.}
  \label{tab:breadth}
  \small
  \begin{tabular}{@{}p{3.7cm}p{0.55cm}p{0.55cm}p{0.55cm}p{0.55cm}p{0.55cm}p{0.55cm}p{1.1cm}p{1.25cm}@{}}
    \toprule
    Report & R1A & R1B & R1C & R2A & R2B & R2C & Best GR & Best Naive \\
    \midrule
    FancyBear / APT28 LayeredMesh & 71 & 76 & 83 & 43 & 16 & 68 & 76 & 83 \\
    LABYRINTH CHOLLIMA TxRLoader & --- & --- & --- & 67 & 70 & 48 & 70 & 48 \\
    Cyble OpSindoor (APT36) & --- & --- & --- & 0* & 16 & 61 & 16 & 61 \\
    LABYRINTH CHOLLIMA macOS & 62 & 16 & 0 & --- & --- & --- & 62 & 0 \\
    Vishing / Help-desk MSC & 77 & 16 & 75 & 76 & 16 & 80 & 77 & 80 \\
    Zimbra LFI & 69 & 60 & 61 & 81 & 16 & 68 & 81 & 68 \\
    Renegade Jackal / Micropsia & 0* & 70 & 78 & 80 & 16 & 70 & 80 & 78 \\
    LAPSUS\$ insider recruitment & 60 & 16 & 80 & --- & --- & --- & 60 & 80 \\
    BADBOX2 Android backdoor & 75 & 71 & 66 & 0* & 71 & 80 & 75 & 80 \\
    \bottomrule
  \end{tabular}
\end{table*}

\begin{figure*}[t]
  \centering
  \includegraphics[width=0.6\textwidth]{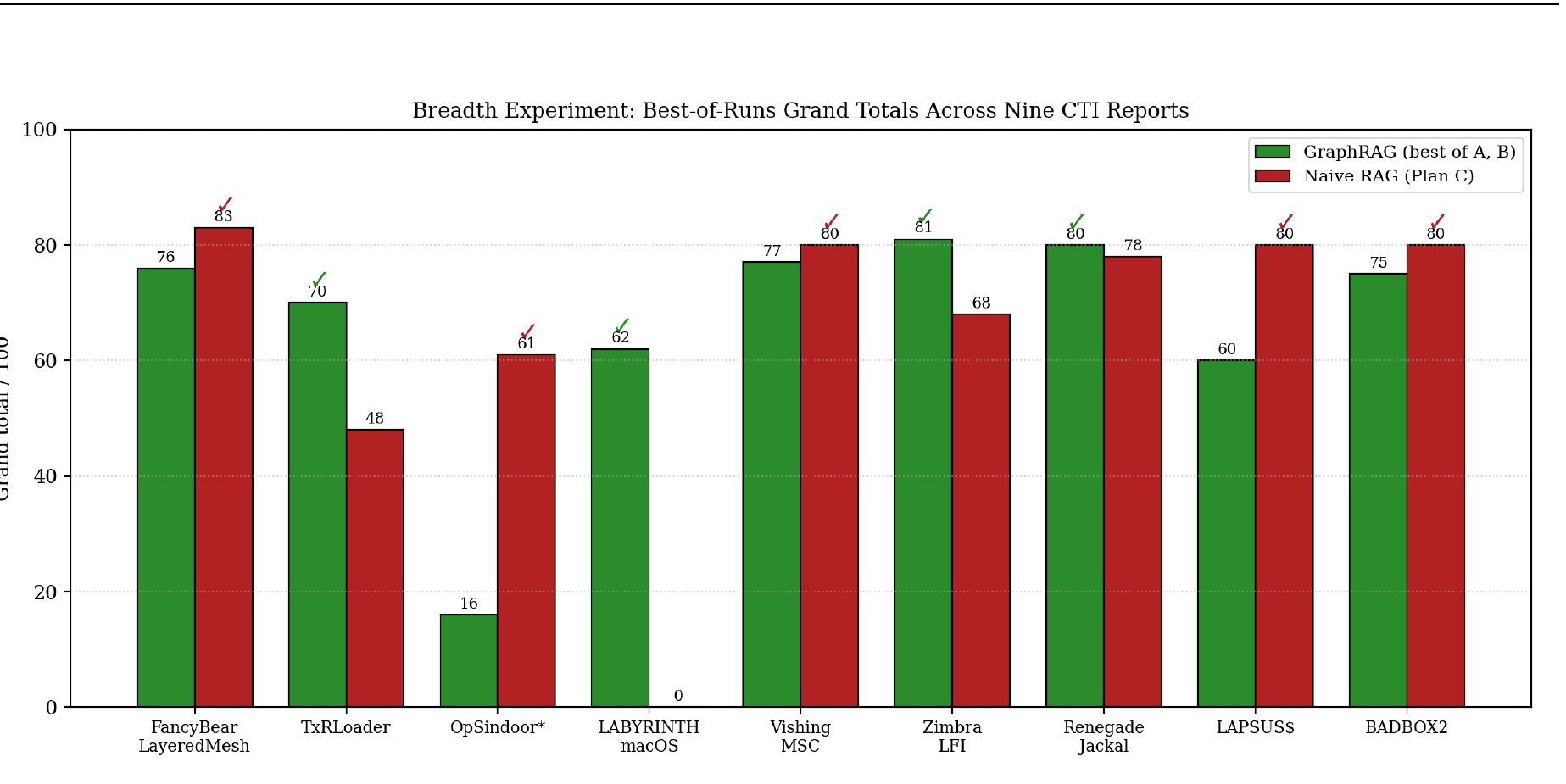}
  \caption{Best-of-runs grand totals across all nine CTI reports. GraphRAG $= \max(A,B)$ over both runs; Naive $=$ best Plan C run. Check marks mark the winner.}
  \Description{A bar chart comparing best-of-runs grand totals for GraphRAG (green) versus Naive RAG (red) across nine CTI reports. GraphRAG wins on TxRLoader (70 vs 48), LABYRINTH macOS (62 vs 0), Zimbra LFI (81 vs 68) and Renegade Jackal (80 vs 78). Naive RAG wins on FancyBear (76 vs 83), OpSindoor (16 vs 61), Vishing (77 vs 80), LAPSUS dollar (60 vs 80) and BADBOX2 (75 vs 80).}
  \label{fig:breadth}
\end{figure*}

\textbf{Win count on grand totals.} Taking the best-of-runs maximum for each pipeline, GraphRAG wins four of nine reports (TxRLoader, LABYRINTH macOS, Zimbra, Renegade Jackal) and Naive RAG wins five (FancyBear, OpSindoor, Vishing, LAPSUS\$, BADBOX2). Three of these results deserve qualification. First, the OpSindoor 16-vs-61 outcome is a GraphRAG silent failure: Plan B fell back to its 16/100 sparse-graph floor and Plan A produced a judge-parse 0; the value 16 is the rubric floor, not a genuine plan-quality measurement. Second, the FancyBear and Vishing margins are within four points and are within the noise of a single 8B-judge run on long input. Third, the LAPSUS\$ report is two pages, below the GraphRAG community-detection threshold, which biases the architectural comparison against GraphRAG by construction. Net of these caveats, the breadth experiment is approximately tied on grand totals.

\textbf{Pyramid composition: the discriminating story.} When the underlying detection composition is examined rather than the grand total, the GraphRAG-versus-Naive-RAG comparison stops being tied. Plan A reaches the host-artefact-or-higher tier (L5+) on three of seven Run 1 reports and the TTP tier (L7) on Vishing; Plan C reaches L5 on zero reports and L7 only on FancyBear, a report uncommonly rich in tradecraft prose. The plans Naive RAG produces at a nominal L4 dominant level still carry a long fragile tail at L1--L3: 45\% of Plan C detections in the V2 deep-dive sit at L1--L3, versus only 15\% of Plan A detections. The single judge criterion that captures this composition, detection\_durability, separates the pipelines cleanly: Plan A retains 100\% on the deep-dive, Plan C retains 29\%. Because Tier 2 is what decides whether the plan is still firing 48 hours after the report becomes public, the discriminating metric for an SOC choosing a pipeline is Tier 2, not the grand total. On Tier 2 and on the IOC rotation survival data of Table~\ref{tab:rotation}, GraphRAG is the superior approach.

\subsection{Failure Modes Observed}
\textbf{(i) PDF extraction failures (Run 1 only).} Two reports --- TxRLoader and OpSindoor --- failed GraphRAG indexing in Run 1 because vertically rendered watermarks (the Cyble logo letters, the EclecticIQ partner banner) fragmented into one-character lines that the entity extractor treated as candidate entities. The \texttt{post\_clean()} routine introduced for Run 2 eliminates this and recovered both reports.

\textbf{(ii) GraphRAG Global silent failure on sparse graphs.} Plan B returns the 16/100 rubric floor when the level-2 community structure is too thin, typically reports under five pages or with low entity density (five of seven Run 2 reports; the LABYRINTH macOS and LAPSUS\$ Run 2 cell-failures are extreme cases of the same mode).

\textbf{(iii) Judge JSON parsing failure.} Three Plan A cells (Run 2 OpSindoor, Run 2 BADBOX2, Run 1 Renegade Jackal) are 0/100 because the 8B judge emitted malformed JSON; the plans themselves were generated successfully, and manual inspection confirms they are comparable in quality to the corresponding non-failed runs. The failure is in the judge, not the GraphRAG pipeline; substituting a 13B-class judge would eliminate this class and mechanically lift GraphRAG's win count.

\section{Discussion}

\subsection{Why Each Pipeline Lands Where It Does on the Pyramid}
The three pipelines occupy three different Pyramid levels for three different and architecturally explicable reasons. \textbf{GraphRAG Local reaches L7-TTP} because L7 detections describe behavioural chains. A representative chain: Outlook spawns \texttt{mshta.exe}, which spawns \texttt{powershell.exe} with an HTA payload, followed within thirty seconds by an outbound TLS connection to port 8080 with a specific JA3 hash. Producing such a detection requires the generation model to know about four entities and the relationships among them; Local Search retrieves precisely this multi-entity neighbourhood, starting from the seed entities and pulling in their direct relationships, the text units describing those relationships, and the community summary that names the chain. The L7 detection is then a near-direct verbalisation of the retrieved subgraph.

\textbf{GraphRAG Global underperforms on sparse reports} because Global Search is designed for corpus-wide sense-making queries over many documents. When the corpus is a single CTI report the level-2 community structure is typically two to four communities, below the threshold at which map-reduce summarisation produces a meaningful answer, producing the 16/100 silent-fail mode. The mode is reproducible and well understood; the practical mitigation --- unimplemented in this work --- is to fall back to Local Search when Global Search returns less than a threshold output length.

\textbf{Naive RAG clusters at L4} because the text spans most likely to be retrieved by cosine similarity for a threat-hunting query are the spans that describe network observables. CTI reports are generally written network-first (fast-flux DNS, 60-second beacons, Cloudflare-fronted domains), and L4 is the layer at which network observables become detection-grade. The retrieval is doing exactly what it is supposed to do; the resulting plan is exactly as fragile as the source prose. The eight chunks most similar to the query also tend to be the chunks that name the loudest entities (C2 IPs, file hashes, actor names), which produces the long L1--L3 fragile tail observed in Plan C across the breadth experiment.

\subsection{Operational Implications for SOC Teams}
Our findings have the following three operational implications. \textbf{(1) Deploy GraphRAG Local as the primary retrieval back-end.} It is the only mode in this work whose plans reach dominant L7-TTP on multi-page CTI reports, and L7-TTP detections are the ones that survive the time window between the SOC reading the report and the adversary rotating infrastructure. \textbf{(2) Deploy GraphRAG Global as a secondary mode with an output-length fallback.} When Global succeeds it produces strong plans; when it silently fails the pipeline must detect the 75-character error fallback and route to Local. A simple length threshold on the Global output is sufficient. \textbf{(3) Retain Naive RAG only for fact-lookup queries.} Naive RAG is appropriate for pointed questions (``what IPs did this campaign use?'') but produces threat-hunting plans whose useful life is measured in hours after the report becomes public knowledge.

The data also delivers a methodological lesson independent of the GraphRAG-versus-Naive comparison. The single largest movement in plan quality observed in this work was the V1-to-V2 prompt change, not the retrieval-back-end change. Detection-engineering generation is acutely sensitive to the contract the prompt imposes; the hardened prompt's mandatory Pyramid tagging, mandatory field-level queries, and explicit IOC-rotation reasoning section are doing most of the work, and the retrieval back-end then determines the ceiling.

\subsection{Threats to Validity and Limitations}
Four caveats apply. First, the deep-dive grand-total margin (80 vs 78) is too small to be statistically defensible from a single judge run; the headline finding rests on the Tier 2 composition (Table~\ref{tab:rotation}) and the breadth experiment, not the grand total. Second, the breadth experiment's 8B judge has JSON-emission stability at the edge of what is reliable on long input; two judge-parsing failures in the seven-report batch is a meaningful error rate. Third, the GraphRAG Global silent-failure mode appears on a majority of the breadth corpus and partially confounds the GraphRAG-as-a-family comparison; reporting GraphRAG as $\max(A,B)$ is conservative, but a fallback-aware system would shift more reports into the GraphRAG-wins column. Fourth, the corpus is weighted toward nation-state APT reports from four vendors; findings should be re-validated on commodity malware advisories and internal telemetry-driven hunt requests.

\section{Conclusions and Future Work}
This project asked whether a knowledge-graph-aware retrieval back-end produces threat hunting plans that are materially less IOC-fragile than a Naive vector RAG, when both are driven by the same generation model and the same hardened generation prompt. The evidence supports the hypothesis on the discriminating metrics --- Pyramid-of-Pain composition and IOC-rotation survival --- even though grand-total scores across the breadth experiment are approximately tied. On the APT28 deep-dive, GraphRAG Local plans retain 100\% of their detection surface after a complete IOC rotation; Naive RAG plans retain 29\%. On the full nine-report breadth experiment, GraphRAG reaches the durable L4--L7 layer on every successful report and reaches L5 or higher on three of seven Run 1 reports; Naive RAG clusters at L4 with a long fragile L1--L3 tail and reaches L5 or higher on zero reports excluding the FancyBear special case. Net of three judge-parsing failures and two short-report community-detection failures, the breadth experiment is consistent with the deep-dive: GraphRAG is the architecturally superior approach for IOC-resilient threat hunting plan generation.

Three directions follow for future work: (i) an order-of-magnitude scale-up to fifty to one hundred reports across a wider vendor base, with statistical testing on per-criterion deltas; (ii) replacement of the 8B judge with a 13B-class cybersecurity judge to eliminate the JSON-parsing failure mode and tighten the scoring distribution; and (iii) two engineering improvements surfaced by the breadth experiment --- a Local-Search fallback when GraphRAG Global returns too little output, and an LLM-based plan-quality precheck that runs before the judge to catch obviously degenerate outputs. With these in place, a fully autonomous CTI-to-detection pipeline that is trustworthy enough for production SOC deployment is within reach.

\section*{Ethics and Privacy Statement}
This work processes only publicly published vendor CTI advisories and one open-source PDF advisory; no private, proprietary, or personally identifiable data was used. All experiments ran on a single local workstation with no external API calls, so no report content was transmitted to a third party during indexing, generation, or judging. The dual-use risk is that the same retrieval-and-generation pipeline used to draft defensive hunting queries could in principle be redirected to draft offensive tooling from a CTI report; the outputs studied here are detection queries (Splunk SPL, Sentinel KQL) rather than exploit or intrusion code, which limits but does not eliminate this risk. Because the judge and generation models are both small, locally-hosted models operating at the edge of their reliable capability on long, technical input, plans produced by this pipeline should be reviewed by a human detection engineer before deployment rather than pushed directly into production SOC tooling.

\appendix

\section{V2 Hardened Prompt --- Non-Negotiable Rules}
\label{app:prompt}
The V2 hardened generation prompt prepends five non-negotiable rules to the seven-section plan template inherited from V1.

\begin{enumerate}
\item \textbf{Source fidelity.} Every IOC listed in the plan must appear verbatim in the source intelligence and be tagged \texttt{[SRC]} so that its origin is unambiguous. Invented IPs, domains, hashes or ATT\&CK IDs are explicitly forbidden; the prompt instructs the model to omit a candidate detection rather than back-fill it with a plausible but unverifiable indicator.

\item \textbf{Query executability.} Every detection query must include real, telemetry-grade field names. The prompt enforces this with an explicit BAD example, the prose string ``search for unusual PowerShell execution,'' and an explicit GOOD example, a six-line Splunk SPL query: \textit{index=endpoint sourcetype=sysmon EventCode=1 parent\_image=*\textbackslash powershell.exe NOT (Image=*\textbackslash conhost.exe) | stats count by Image, CommandLine, host}. Anything closer to the BAD example than the GOOD example is rejected by the rubric's detection\_readiness ceiling.

\item \textbf{Pyramid tagging.} Every detection in the plan must be tagged with its Pyramid level using a shorthand: \textit{[L1-Hash]}, \textit{[L2-IP]}, \textit{[L3-Domain]}, \textit{[L4-NetworkArtifact]}, \textit{[L5-HostArtifact]}, \textit{[L6-Tool]}, \textit{[L7-TTP]}. The prompt also tells the model that L1--L3 detections expire within forty-eight hours of report publication while L4--L7 detections survive adversary infrastructure rotation, giving the model the operational context for its own tagging.

\item \textbf{ATT\&CK depth.} Every ATT\&CK ID cited in the plan must be paired with three artefacts: an observed-evidence quote from the source report, a field-level detection query, and a named false-positive baseline. ATT\&CK IDs listed without attached detection logic are explicitly called out by the prompt as ``noise'' and are penalised by the attck\_coverage ceiling at the rubric layer.

\item \textbf{Behavioural priority.} The prompt enforces the structural ordering of the plan. Section 2 of the output is the behavioural detection chains and is labelled \textsc{Primary} because these detections survive IOC rotation. Section 3 of the output is the IOC hunting list and is labelled \textsc{Fragile} with explicit TTL annotations (approximately 48 hours for IPs and domains, approximately 30 days for hashes). The ordering is what allows an SOC L2 analyst to deploy the durable detections first and treat the IOC list as enrichment rather than primary signal.
\end{enumerate}

The five rules together change the V2 prompt from a generic threat-hunting template into a contract: every detection it produces is tagged, queryable, source-grounded, ATT\&CK-justified and ordered by durability. The empirical effect of this contract on the APT28 deep-dive --- a fifty-five to seventy-four point grand-total swing across all three retrieval back-ends --- is reported in Section~\ref{sec:v1v2}.

\begingroup
\small
\bibliographystyle{ACM-Reference-Format}
\bibliography{refs}
\endgroup

\end{document}